\documentclass{JHEP3}
\usepackage{amssymb}
\usepackage{amsmath}
\usepackage{epsfig,multicol,bbm}

\abstract{We study the three-dimensional gravity with torsion given by the Mielke-Baekler (MB) model coupled to gravitational Chern-Simons term, and that possess electric charge described by Maxwell-Chern-Simons electrodynamics. We find and discuss this theory's charged black holes solutions and uncharged solutions. We find that for vanishing torsion our solutions by means of a coordinate transformation can be written as three-dimensional Chern-Simons black holes. We also discuss a special case of this theory, Topologically Massive Gravity (TMG) at chiral point, and we show that the logarithmic solution of TMG is also a solution of the MB model at a fixed point in the space of parameters. Furthermore, we show that our solutions generalize G\"{o}del type solutions in a particular case. Also, we recover BTZ black hole in Riemann-Cartan spacetime for vanishing charge.}
\maketitle 

\title{Exact solutions in 3D gravity with torsion}
\author{P.A. Gonz\'{a}lez$^{a,b}$, Yerko V\'{a}squez$^{c}$ \\

$^a$Escuela de Ingenier\'{i}a Civil en Obras Civiles. Facultad de Ciencias F\'{i}sicas y Matem\'{a}ticas, Universidad Central de Chile, Avenida Santa Isabel 1186, Santiago, Chile.\\

$^b$Universidad Diego Portales, Casilla 298-V, Santiago, Chile.\\

$^c$Departamento de Ciencias F\'{\i}sicas, Facultad de Ingenier\'{\i}a,
Ciencias y Administraci\'{o}n, Universidad de La Frontera, Avenida Francisco
Salazar 01145, Casilla 54-D, Temuco, Chile.\\

E-mail:
\email{pgonzalezm@ucentral.cl}, \email{yvasquez@ufro.cl}.}

\begin{document}
\section{Introduction}
In recent years, there has been a remarkable activity in the study of three-dimensional models of gravity.
In particular, we have the BTZ black hole~\cite{BTZ},
which is a solution to the Einstein equations with a negative cosmological
constant. This black hole solution has interesting both classical and
quantum properties, and it shares several features of the Kerr black hole of
four-dimensional General Relativity (GR)~\cite{Carlip}. In fact, the
existence of BTZ black holes is what makes the three-dimensional gravity a
striking toy model.

Recently, remarkable attention was addressed to Topologically Massive
Gravity (TMG), a generalization of
three-dimensional GR that amounts to augment the Einstein-Hilbert action adding a Chern-Simons gravitational term,~\cite{deser}. Here, the propagating degree of freedom is a massive graviton. TMG also admits the BTZ (and other)
black holes as exact solutions. The renewed interest on TMG relies on the
possibility of constructing a chiral theory of gravity at a special point of
the space of parameters, what it was suggested in Ref.~\cite{Li:2008dq}. This
idea has been extensively analyzed in the last three years~\cite{Strominger:2008dp}-\cite{Henneaux:2009pw} and
it gave raise a fruitful discussion that ultimately led to a much better
understanding of the model~\cite{Maloney:2009ck}. Also, recently has been
showed that in three-dimensional massive gravity, where the action is given
by the Einstein-Hilbert action having square-curvature terms which gives
raise to field equations with a second order trace admits exacts Lifshitz
metrics and black hole solutions, which are asymptotically Lifshitz,~\cite{AyonBeato:2009nh}. However, the formulation of a quantum theory of gravity
is an open problem in theoretical physics. Some approach at this problem has
been done in superstring theory and loop quantum gravity. Thus, it is worth
to explore new possibilities such as gravity with torsion which in (2+1)D
are relationships with the continuum theory of lattice defects in solid
physics,~\cite{Tresguerres:1992ew, Kawai:1994qw}.

So, a possible extension of TMG is to add torsion in the game. In fact, the three-dimensional theory of gravity
that includes torsion (along with the Einstein-Hilbert and the
gravitational Chern-Simons term) is known with the name of Mielke-Baekler
theory~\cite{Baekler}, and this model is itself a Chern-Simons theory which
includes a translational term as well, and therefore this is a striking
mathematical model in its own right. Moreover, in regarding the applications to
physics, it is well-known that the introduction of torsion in the game often
induces new physical effects and changes the local degrees of freedom of the
theory. In Ref.~\cite{Klemm} Klemm and Tagliabue investigated the introduction of
torsion in three dimensions and discussed it, according to AdS/CFT
correspondence. This is very interesting if, in addition, one takes into
account that a generalization of the BTZ black hole solution with torsion
also exists in the literature~\cite{garcia}. In a more recent context, and also
related to AdS/CFT applications, the so-called degenerate point of the
Mielke-Baekler theory (for which the equations of motion associated to the
vielbein coincides with those associated to the spin connection) can be
shown to exhibit features that are similar to the so-called chiral point of
TMG~\cite{Gaston} (e.g. leading to the same value of the boundary central
charges~\cite{Klemm}). Three-dimensional gravity with torsion was also
recently considered in Ref.~\cite{Troncoso}, where the supersymmetric extension
in the Chern-Simons formulation was investigated. Exact solutions with
torsion in three dimensions were analyzed recently in Refs.~\cite{Blagojevic}-\cite{Blagojevic:2009xw}.

In this paper, we study a general 3D model of gravity based on Riemann-Cartan Geometry, whose fundamental fields are both metric and torsion. We consider topological model of gravity proposed by Mielke and Baekler~\cite{Baekler}, and we include the matter content given by the Maxwell action
augmented by a topological Chern-Simons term, that is, Einstein-Cartan gravity with arbitrary cosmological constant coupled to a gravitational
topological term and topologically massive electrodynamics. Chern-Simons modifications to gravity were first considered in 2+1
dimensions~\cite{deser}. These modifications could in principle arise as
a truncation of a consistent theory of quantum gravity, such as string
theory or M-theory. A discussion on how the gravitational Chern-Simons term
may appear from compactifying $R^{4}$-terms of 11D supergravity has recently
been given in~\cite{Santamaria:2011cz}. On the other hand, in what concerns the translational Chern-Simons
term, for example, many models of loop quantum gravity or string theory
predict a coupling/term in the action with the Nieh-Yan topological
invariant which is associated to it; the inverse of the coupling constant of
such term is often called the "Immirzi parameter". In turn, it is perfectly
conceivable that both the gravitational as the translational Chern-Simons
gravitational terms appear as reductions of fundamental theories. Also, it
is known that Einstein-Maxwell-Chern-Simons theory in (2+1) dimensions can
be viewed as a lower dimensional model for the bosonic sector of 5-dimensional
supergravity~\cite{Cremmer}. In references~\cite{Campbell:1990fu} and~\cite{Campbell:1990ai} the authors investigated the structure of the
gravitational Chern-Simons term modifications to General Relativity, and
also showed how such terms could arise as a low-energy consequence of string
theory; in~\cite{Alexander:2009tp} it is discussed that these terms could also
arise by anomaly cancelation in particle physics, string theory and loop
quantum gravity.

Independent gravitational fields in this framework are the vielbein $e$ and
the Cartan spin connection $\omega $. On the other hand, the curvature and torsion
describe the gravitational dynamics. We write an action for this theory using first order
formalism of Poincar\'{e} gauge theory of gravity (PGTG). Then, the theory in second order formulation, presented in terms of the metric field $g$ and affine connection $\Gamma$, is easily obtained using the
metricity condition $Dg=0$, with $D$ being a covariant
derivative with respect to the affine connection. This method allows us
to find new analytic solutions to this theory representing charged
rotating black hole solutions that, in turn, generalize previous solutions
reported in the literature. For vanishing torsion, these solutions by means of a
coordinate transformation can be written as a three-dimensional Chern-Simons
black hole. Also, we show that the logarithmic solution of TMG, found in Ref.~\cite{Garbarz:2008qn}, is solution of the Mielke-Baekler model at a special point of space parameters.
We will also briefly comment a solution that generalizes the G\"{o}del spacetime
obtained in Ref.~\cite{Banados} in terms of spacetime that includes torsion and gravitational
topological term. The G\"{o}del like solution is supported by an abelian gauge
field, and it is necessary to include an additional Chern-Simons term which produces the energy-momentum tensor of a pressureless perfect fluid. In Ref.~\cite{Banados} it is show that in $2+1$ dimensions the Maxwell field minimally
coupled to gravity can be the source of such a fluid, when it is increased
with a topological mass $\mu _{E}$.

In the case of pure gravity, knowing the black
hole content in the spectrum turns out to be a crucial point to fully
understand the theory's properties. In particular, in the asymptotically AdS$_{3}$ sector this is important to try to reconstruct the dual CFT$_{2}$.
The appropriate classification of bulk geometries that contributes to
the partition function is a crucial step that could shed light on the
quantum gravity theory. Of course, the problem of quantizing the theory in
presence of matter (e.g. of U(1) matter) is far from being a tractable task
to our knowledge. Nevertheless, the question about the theory's black hole content still represents a well motivated classical problem.

The structure of the paper is as follows. In section $2$ we introduce the
model and we derive the field equations. Then, in section $3$ we solve the
field equations with an ansatz for the vielbein and the gauge field, and in
section $4$ we analyze the solution for some particular cases and
limits. To conclude, our remarks are in section $5$.

\section{Three-dimensional gravity in Riemann-Cartan space}
In the Einstein-Cartan geometry the basic gravitational fields are the
vielbein 1-form $e^{a}=e_{\mu}^{a}dx^{\mu}$ and the spin connection 1-form
$\omega^{ab}=\omega _{\mu}^{ab}dx^{\mu}$\footnote{%
Latin indices label the components with respect to a local Lorentz frame and
Greek indices refers to the coordinate frame.}, we take our local coordinates to be $x^{\mu }=t,r,\phi$.
By simplicity in the notation, it is standard to work with a dual spin
connection
\begin{equation}
\omega ^{a}=-\frac{1}{2}\epsilon ^{abc}\omega _{bc}~.
\end{equation}%
So, the torsion $T^{a}$ and the curvature $R^{a}$ are given by
\begin{equation}
T^{a}=de^{a}+\epsilon_{bc}^{a}\omega^{b}e^{c}~,
\end{equation}%
\begin{equation}
R^{a}=d\omega^{a}+\frac{1}{2}\epsilon_{bc}^{a}\omega^{b}\omega ^{c}~.
\label{Curvature}
\end{equation}

We will consider the model proposed by Mielke-Baekler charged under
Chern-Simons electrodynamics, which is given by the following action
\begin{equation}
S=\int \left( \frac{1}{\kappa }e_{a}R^{a}-\frac{\Lambda }{3}\epsilon
_{abc}e^{a}e^{b}e^{c}+\alpha _{3}\left( \omega ^{a}d\omega _{a}+\frac{1}{3}
\epsilon _{abc}\omega ^{a}\omega ^{b}\omega ^{c}\right) +\alpha
_{4}e_{a}T^{a}+\mathcal{L}_{F}\right) ~,  \label{action}
\end{equation}%
where
\begin{equation}
\mathcal{L}_{F}=-\frac{1}{2}F^{\ast }F-\frac{\mu _{E}}{2}AF~,
\label{Mlagrangian}
\end{equation}%
$\kappa =8\pi G$, $\alpha _{3}$ and $\mu _{E}$ are the gravitational and
electromagnetic Chern-Simons coupling constants, respectively, $\alpha _{4}$
is the torsion coupling, $\Lambda $ is a cosmological constant, $A$ is the potential 1-form, $F=dA$
and $\epsilon _{abc}$ is the completely antisymmetric symbol with $\epsilon _{012}=+1$. Using the metricity condition we arrive to the following
useful relation
\begin{equation}
\omega ^{a}=\bar{\omega}^{a}+K^{a}~,  \label{conecction}
\end{equation}%
where $\bar{\omega}^{a}$ is the Riemannian connection and $K^{a}$ is the
contortion 1-form. The contortion is defined through
\begin{equation}
T^{a}=\epsilon _{bc}^{a}K^{b}e^{c}~.  \label{contortion}
\end{equation}%
Relation~(\ref{conecction}) allows us to express the Cartan curvature in terms of the
Riemannian curvature and the contortion as
\begin{equation}
R^{a}=\bar{R}^{a}+\bar{D}K^{a}+\frac{1}{2}\epsilon _{bc}^{a}K^{b}K^{c}~.
\label{curv2}
\end{equation}
Where, $\bar{D}$ is the covariant derivative with respect to $\bar{\omega}^{a}$. The spacetime metric is $g_{\mu\nu}=\eta_{ab}e_{\mu}^{a}e_{\nu}^{b}$
and we have adopted the convention $\eta _{ab}=diag(+1,-1,-1)$.

In order to obtain the field equations we vary the action with respect to the
independent fields $e^{a}$ and $\omega ^{a}$ getting
\begin{equation}
\frac{1}{\kappa }R^{a}-\Lambda \epsilon _{bc}^{a}e^{b}e^{c}+2\alpha
_{4}T^{a}=-\frac{\delta \mathcal{L}_{F}}{\delta e^{a}}~,
\end{equation}
\begin{equation}
2\alpha _{3}R^{a}+\alpha _{4}\epsilon _{bc}^{a}e^{b}e^{c}+\frac{1}{\kappa }%
T^{a}=0~,
\end{equation}
from these expressions we obtain the torsion and the curvature, which can be expressed as
\begin{equation}
T^{a}-\frac{a}{2}\epsilon _{bc}^{a}e^{b}e^{c}=\frac{m}{2}\chi ^{a}~,
\label{eom1}
\end{equation}%
\begin{equation}
R^{a}-\frac{b}{2}\epsilon _{bc}^{a}e^{b}e^{c}=-\frac{n}{2}\chi ^{a}~,
\label{eom2}
\end{equation}%
where
\begin{equation}
\chi _{a}=-\frac{\delta \mathcal{L}_{F}}{\delta e^{a}}~,  \label{current}
\end{equation}%
is the (Maxwell) energy-momentum current and we have introduced the
following constants
\begin{equation}
a=\frac{\alpha _{3}\Lambda +\frac{\alpha _{4}}{2\kappa }}{\alpha _{3}\alpha
_{4}-\left( \frac{1}{2\kappa }\right) ^{2}}~,\quad b=-\frac{\alpha _{4}^{2}+%
\frac{1}{2\kappa }\Lambda }{\alpha _{3}\alpha _{4}-\left( \frac{1}{2\kappa }%
\right) ^{2}}~,
\end{equation}%
\begin{equation}
\quad m=\frac{\alpha _{3}}{\alpha _{3}\alpha _{4}-\left(
\frac{1}{2\kappa }\right) ^{2}}~,\quad n=\frac{\frac{1}{2\kappa }}{\alpha
_{3}\alpha _{4}-\left( \frac{1}{2\kappa }\right) ^{2}}~,  \label{const}
\end{equation}%
with the condition $\alpha _{3}\alpha _{4}-\left( \frac{1}{%
2\kappa }\right) ^{2}\neq 0$~.
The energy-momentum tensor is calculated by
\begin{equation}
S_{b}^{a}=^{\ast }(e^{a}\chi _{b})=-F^{ac}F_{bc}+\frac{1}{4}\delta
_{b}^{a}(F^{cd}F_{cd})~,  \label{energy momentun}
\end{equation}%
and the energy-momentum current can be written as
\begin{equation}
\chi _{a}=\epsilon _{abc}s^{b}e^{c}~,
\end{equation}%
with%
\begin{equation}
s^{a}=-\left( S_{b}^{a}-\frac{1}{2}\delta _{b}^{a}S\right) e^{b}~,  \label{s}
\end{equation}%
where $S=S_{a}^{a}$, is the trace of the energy-momentum tensor. Replacing,
the expression for the current, in the gravitational equations, we obtain
\begin{equation}
T^{a}=\frac{1}{2}\epsilon _{bc}^{a}(ae^{b}+ms^{b})e^{c}~,  \label{eom3}
\end{equation}%
\begin{equation}
R^{a}=\frac{1}{2}\epsilon _{bc}^{a}(be^{b}-ns^{b})e^{c}~,  \label{eom4}
\end{equation}%
and by using Eq.~(\ref{eom3}) we find that the contortion can be written as
\begin{equation}\label{relationk}
K^{b}=\frac{1}{2}\left( ae^{b}+ms^{b}\right) ~.
\end{equation}

Now, by varying the action with respect to 1-form $A$ we obtain the modified
Maxwell equations
\begin{equation}
d^{\ast }F+\mu _{E}F=0~,  \label{maxwell}
\end{equation}%
which includes the contribution of the electromagnetic Chern-Simons term. Later by replacing
the expression~(\ref{relationk}) for the contortion in Eq.~(\ref{curv2}), we find the
following expression to the Cartan curvature
\begin{equation}
R^{a}=\bar{R}^{a}+\frac{m}{2}\bar{D}s^{a}+\frac{1}{8}\epsilon
_{bc}^{a}\left( a^{2}e^{b}e^{c}+2ams^{b}e^{c}+m^{2}s^{b}s^{c}\right) ~.
\label{curv3}
\end{equation}%
Taken this last equation together with the second gravitational Eq.~(\ref{eom4}), we finally obtain 
\begin{equation}
\bar{R}^{a}=-\frac{\gamma }{2}\epsilon _{bc}^{a}s^{b}e^{c}-\frac{m}{2}\bar{D}%
s^{a}-\frac{m^{2}}{8}\epsilon _{bc}^{a}s^{b}s^{c}+\frac{1}{2}\Lambda
_{eff}\epsilon _{bc}^{a}e^{b}e^{c}~,  \label{curv}
\end{equation}%
where we have defined
\begin{equation}
\gamma =n+\frac{am}{2}~,
\end{equation}%
the effective cosmological constant as
\begin{equation}
\Lambda _{eff}=b-\frac{a^{2}}{4}=-\frac{1}{l^{2}}~,  \label{lamda}
\end{equation}%
and
\begin{equation}
\bar{D}s^{a}=ds^{a}+\epsilon _{bc}^{a}\bar{\omega}^{b}s^{c}~,
\end{equation}%
as the covariant derivative (with respect to $\bar{\omega}$) of $s^{a}$.

\section{Solutions}
Now, for the metric we consider the following stationary circularly symmetric ansatz \footnote{%
Note that this form for the metric agrees with the one used in \cite%
{Gurses}, see Eq.~(39) therein and consider $r=\bar{r}^{2}$.}
\begin{equation}
ds^{2}=\frac{\psi (r)}{f(r)}dt^{2}-f(r)\left( d\phi +\frac{Cr}{f(r)}%
dt\right) ^{2}-\frac{1}{\psi (r)}dr^{2}~,  \label{metric}
\end{equation}%
in the local coordinates $x^{\mu }=t,r,\phi$, with $0\leq \phi \leq 2\pi$ and the following ansatz for the U(1) field
\begin{equation}
A=A_{t}dt+A_{\phi }d\phi  \label{electric field}
\end{equation}
which is a spherically symmetric gauge field in the gauge $A_{r}=0$. Also,
we consider $A_{t}$ to be a constant and $A_{\phi }=I+Hr$ , with $I$ and $H$
as constants, this yields
\begin{equation}
F=dA=Hdrd\phi~.
\end{equation}
In addition, we will choose $f(r)$ and $\psi(r)$ as quadratic functions of $r$
\begin{equation}
f(r)=Mr^{2}+Nr+L~,
\end{equation}%
\begin{equation}
\psi (r)=Pr^{2}+Qr+R~,
\end{equation}
where $M,N,L,P,Q,R$ are integration constants to be determined.

The orthonormal basis is determined up to a local Lorentz transformation and
we choose this basis to be
\begin{equation}
e^{0}=\sqrt{\frac{\psi (r)}{f(r)}}dt~,  \label{E0}
\end{equation}%
\begin{equation}
e^{1}=\frac{1}{\sqrt{\psi (r)}}dr~,  \label{E1}
\end{equation}%
\begin{equation}
e^{2}=\sqrt{f(r)}\left( d\phi +\frac{Cr}{f(r)}dt\right) ~,  \label{E2}
\end{equation}%
and for the electric field
\begin{equation}
F=E(r)e^{0}e^{1}-B(r)e^{1}e^{2}~,  \label{F}
\end{equation}%
by contracting the electromagnetic tensor with itself we obtain
 the electromagnetic invariant
\begin{equation}
F_{ab}F^{ab}=2(B^{2}-E^{2})~,
\end{equation}%
from this result the energy-momentum tensor reads
\begin{equation}
S_{b}^{a}=\left(
\begin{array}{ccc}
\frac{1}{2}(E^{2}+B^{2}) & 0 & EB \\
0 & \frac{1}{2}(E^{2}-B^{2}) & 0 \\
-EB & 0 & -\frac{1}{2}(E^{2}+B^{2})%
\end{array}%
\right) ~,  \label{stress}
\end{equation}%
and the trace of the Maxwell energy-momentum tensor is
\begin{equation}
S=\frac{1}{2}(E^{2}-B^{2})~,  \label{trace2}
\end{equation}%
from the modified Maxwell equations we get the following pair of equations
\begin{equation}
\sqrt{\psi }E^{\prime }+\frac{1}{2}\sqrt{\psi }\frac{f^{\prime }}{f}E+\mu
_{E}B=0~,  \label{maxwell1}
\end{equation}%
\begin{equation}
\sqrt{\psi }B^{\prime }+\frac{1}{2\sqrt{\psi }}\left( \psi ^{\prime }-\psi
\frac{f^{\prime }}{f}\right) B+\left( C-Cr\frac{f^{\prime }}{f}+\mu
_{E}\right) E=0~.  \label{maxwell2}
\end{equation}

Taking into account the ansatz for the electric field in local coordinates
is $F=Hdrd\phi $, we obtain the following electromagnetic field that
generate the gravitational field
\begin{equation}
E(r)=\frac{CHr}{\sqrt{f}}~,  \label{electric}
\end{equation}%
\begin{equation}
B(r)=-H\sqrt{\frac{\psi }{f}}~,  \label{magnetic}
\end{equation}%
with $C=\mu _{E}$. From the energy-momentum tensor we determine the 1-form $%
s^{a}$ and we get
\begin{equation}
s^{0}=-\frac{1}{4}\left( 3B^{2}+E^{2}\right) e^{0}-EBe^{2}~,  \label{S0}
\end{equation}%
\begin{equation}
s^{1}=-\frac{1}{4}(E^{2}-B^{2})e^{1}~,  \label{S1}
\end{equation}%
\begin{equation}
s^{2}=\frac{1}{4}(B^{2}+3E^{2})e^{2}+EBe^{0}~.  \label{S2}
\end{equation}

The null torsion condition yields the Riemannian connection
\begin{equation}
\bar{D}e^{a}=de^{a}+\epsilon _{bc}^{a}\bar{\omega}^{b}e^{c}=0~,
\end{equation}%
considering the expressions for $e^{a}$ we have
\begin{equation}
\bar{\omega}^{0}=-\frac{1}{2}C\left( 1-r\frac{f^{\prime }}{f}\right) e^{0}-%
\frac{1}{2}\sqrt{\psi }\frac{f^{\prime }}{f}e^{2}~,  \label{W0}
\end{equation}%
\begin{equation}
\bar{\omega}^{1}=-\frac{1}{2}C\left( 1-r\frac{f^{\prime }}{f}\right) e^{1}~,
\label{W1}
\end{equation}%
\begin{equation}
\bar{\omega}^{2}=-\frac{1}{2\sqrt{\psi }}\left( \psi ^{\prime }-\psi \frac{%
f^{\prime }}{f}\right) e^{0}+\frac{1}{2}C\left( 1-r\frac{f^{\prime }}{f}%
\right) e^{2}~,  \label{W2}
\end{equation}%
based on expressions obtained above, we get the Riemannian curvature
\begin{equation}
\bar{R}^{0}=-\frac{1}{2}Cr\frac{f^{\prime \prime }}{f}\sqrt{\psi }%
e^{0}e^{1}-\left( \frac{1}{4}\psi ^{\prime }\frac{f^{\prime }}{f}+\frac{1}{2}%
\psi \frac{f^{\prime \prime }}{f}-\frac{1}{4}\psi \frac{f^{\prime 2}}{f^{2}}+%
\frac{1}{4}C^{2}\left( 1-r\frac{f^{\prime }}{f}\right) ^{2}\right)
e^{1}e^{2}~,  \label{R0}
\end{equation}%
\begin{equation}
\bar{R}^{1}=-\left( \frac{1}{4}\frac{f^{\prime }}{f}\left( \psi ^{\prime
}-\psi \frac{f^{\prime }}{f}\right) +\frac{1}{4}C^{2}\left( 1-r\frac{%
f^{\prime }}{f}\right) ^{2}\right) e^{0}e^{2}~,  \label{R1}
\end{equation}%
\begin{equation}
\bar{R}^{2}=\left( -\frac{1}{2}\psi \frac{f^{\prime \prime }}{f}+\frac{1}{2}%
\psi ^{\prime \prime }-\frac{3}{4}\psi ^{\prime }\frac{f^{\prime }}{f}+\frac{%
3}{4}\psi \frac{f^{\prime 2}}{f^{2}}-\frac{3}{4}C^{2}\left( 1-r\frac{%
f^{\prime }}{f}\right) ^{2}\right) e^{0}e^{1}-\frac{1}{2}Cr\frac{f^{\prime
\prime }}{f}\sqrt{\psi }e^{1}e^{2}~.  \label{R2}
\end{equation}%
Therefore, from the gravitational equations (See appendix A) we obtain the
following algebraic system of equations
\begin{equation}
\frac{m^{2}}{8}\frac{R}{L}H^{4}+\left( \gamma +\frac{3}{2}mC\right)
H^{2}-2M=0~,  \label{ecuacion1}
\end{equation}%
\begin{equation}
3M\frac{R}{L}+C^{2}+4\Lambda _{eff}=\frac{1}{2}\left( \gamma +\frac{5}{2}%
mC\right) H^{2}\frac{R}{L}~,  \label{ecuacion2}
\end{equation}%
with the following constraints between the coefficients
\begin{equation}
P=\frac{R}{L}M+C^{2}~,
\end{equation}%
\begin{equation}
Q=\frac{R}{L}N~,
\end{equation}%
and the function $\psi (r)$ takes the form
\begin{equation}
\psi (r)=\frac{R}{L}f(r)+C^{2}r^{2}~.  \label{phi}
\end{equation}%
Therefore, the solution for gravitational equations is
\begin{equation}
\sigma :=H^{2}\frac{R}{L}=\frac{4}{m^{2}}\left( -\frac{2}{3}\left( \gamma
+m\mu _{E}\right) \pm \sqrt{\frac{4}{9}\left( \gamma +m\mu _{E}\right) ^{2}-%
\frac{1}{3}m^{2}\left( \mu _{E}^{2}+4\Lambda _{eff}\right) }\right) ~,
\label{solution1}
\end{equation}%
\begin{equation}
\rho :=M\frac{R}{L}=-\frac{1}{3}\left( \mu _{E}^{2}+4\Lambda _{eff}\right) +%
\frac{1}{6}\left( \gamma +\frac{5}{2}m\mu _{E}\right) H^{2}\frac{R}{L}~.
\label{solution2}
\end{equation}%

To sum up, Eqs.~(\ref{solution1},~\ref{solution2}) along with the ansatz
Eqs.~(\ref{E0}~-~\ref{F}) and the connection given by Eq.~(\ref{conecction}) represent an exact solution of the theory, (\ref{action}). Next, we will fix $N=2$.

It is worth noticing that for vanishing torsion this solution is related by
a coordinate transformation to analogous solutions in a torsionless theory
of topologically massive gravity coupled to topologically massive
electrodynamics. In fact, the metric given in Eq.~(\ref{metric}) is related to the
metric of Eq.~(3.17) of \cite{Clement2} by the following coordinate transformation
between $r$ and the radial coordinate $\bar{\rho}$ used in that paper

\begin{equation}
\bar{\rho}=-Cr-\frac{RN}{2LC}\frac{1}{1+\frac{MR}{LC^{2}}}. 
\end{equation}

Also, the integration constants in both metrics are related by

\begin{equation}
\beta ^{2}=1+\frac{MR}{LC^{2}},\text{ \ \ \ \ }\rho _{0}^{2}=-\frac{R}{1+%
\frac{MR}{LC^{2}}}+\frac{\frac{R^{2}N^{2}}{4L^{2}C^{2}}}{\left( 1+\frac{MR}{%
LC^{2}}\right) ^{2}}, 
\end{equation}
\begin{equation}
\bar{c}=\frac{M}{C^{2}},\text{ \ \ \ \ \ \ \ \ \ }\omega =-\frac{N}{2C}+%
\frac{RMN}{2LC^{3}}\frac{1}{1+\frac{MR}{LC^{2}}}. 
\end{equation}

Therefore, our solution generalizes the three-dimensional torsionless
Chern-Simons black holes of \cite{Clement2} to spacetime with torsion. Where we have
also included a translational Chern-Simons term in the action. It is
interesting to notice that this topological term induces an effective
cosmological constant even if the bare cosmological constant $\Lambda $ and
the gravitational Chern-Simons coupling $\alpha _{3}$ are both zero (see Eq.~( \ref{lamda})).

\section{Particular cases}
In this section we discuss particular cases of our solution to view the physical content involved by taking some limits.  

\subsection{Case: $\alpha _{3}\alpha _{4}-\left( \frac{1}{%
2\kappa }\right) ^{2}\neq 0$.}
First, let's consider the case $\alpha _{3}=0$ $(m=0$) and $L=0$. In
this case Eqs.~(\ref{ecuacion1}, \ref{ecuacion2}) can be simplified to
\begin{equation}
-\kappa H^{2}=M~,
\end{equation}%
\begin{equation}
3M\frac{Q}{N}+C^{2}+4\Lambda _{eff}=-\kappa H^{2}\frac{Q}{N}~,
\end{equation}%
and we obtain
\begin{equation}
H^{2}\frac{Q}{N}=\frac{1}{2\kappa }\left( \mu _{E}^{2}+4\Lambda
_{eff}\right) ~,
\end{equation}%
\begin{equation}
M\frac{Q}{N}=-\frac{1}{2}\left( \mu _{E}^{2}+4\Lambda _{eff}\right) ~.
\end{equation}%
The conserved charges (See appendix C) can be expressed as
\begin{equation}
\mathcal{Q}=0~,\quad \mathcal{M}=\frac{\left( \mu _{E}^{2}+4\Lambda
_{eff}\right) }{2\kappa }\frac{1}{M}~,\quad \mathcal{J}=0~,
\end{equation}%
where $\mathcal{Q}$, $\mathcal{M}$ and $\mathcal{J}$ denote the electric
charge, the mass and the angular momentum, respectively, and the gauge field is given by
\begin{equation}
A=\frac{\frac{\mu _{E}}{\kappa }}{\sqrt{-\frac{1}{\kappa }M}}dt+\left( \frac{%
\frac{1}{\kappa }}{\sqrt{-\frac{1}{\kappa }M}}+Hr\right) d\phi ~.
\label{potential1}
\end{equation}%
Arriving to the following solution for the metric and gauge field
\begin{equation}
f(r)=\frac{1}{2\kappa l^{2}\mathcal{M}}\left( \mu _{E}^{2}l^{2}-4\right)
r^{2}+2r~,
\end{equation}%
\begin{equation}
\psi (r)=\frac{1}{2l^{2}}\left( \mu _{E}^{2}l^{2}+4\right) r^{2}-2\kappa
\mathcal{M}r~,
\end{equation}%
\begin{equation}\label{potential2}
A=\sqrt{-\frac{1}{2\kappa ^{2}l^{2}\mathcal{M}}\left( \mu
_{E}^{2}l^{2}-4\right) }\left[- \frac{\frac{1}{\kappa }\mu _{E}}{\frac{1}{%
2\kappa ^{2}l^{2}\mathcal{M}}\left( \mu _{E}^{2}l^{2}-4\right) }dt+\left( r-%
\frac{\frac{1}{\kappa }}{\frac{1}{2\kappa ^{2}l^{2}\mathcal{M}}\left( \mu
_{E}^{2}l^{2}-4\right) }\right) d\phi \right] ~.
\end{equation}

The gauge field in Eq.~(\ref{potential2}) is real if
\begin{equation}
-\frac{1}{2\kappa ^{2}l^{2}\mathcal{M}}\left( \mu _{E}^2l^{2}-4\right) >0~.
\end{equation}
Therefore, we have the following cases
\begin{enumerate}
  \item
  [a.]$\mu _{E}^{2}l^{2}-4>0$ and $\mathcal{M}<0$~.
  \item
  [b.]$\mu _{E}^{2}l^{2}-4<0$ and $\mathcal{M}>0$~.
\end{enumerate}
In the first case,
\begin{enumerate}
\item[1.] $f\left( r\right) =0$ at $r=0$ and $r=-%
\frac{4\kappa l^{2}\mathcal{M}}{\mu _{E}^2l^{2}-4}>0$. $f\left( r\right)$ changes sign and there is closed timelike curves (CTC).
\item[2.] $\psi \left( r\right) =0$, at $r=0$ and at $r=\frac{%
4\kappa \mathcal{M}l^{2}}{\mu _{E}^2l^{2}+4}<0$. It's important to note that for $0<r<-\frac{%
4\kappa l^{2}\mathcal{M}}{\mu _{E}^2l^{2}-4}$, Killing vector, $\partial
_{\varphi }$, is spacelike. So, there isn't horizon and the solution represent G\"{o}del Particles.
\end{enumerate}

In the second case,
\begin{enumerate}
\item [1.] $f\left( r\right) =0$ at $r=0$ and $r=-\frac{%
4\kappa l^{2}\mathcal{M}}{\mu _{E}^2l^{2}-4}>0$, therefore, there is CTC.
\item [2.] $\psi \left( r\right) =0$ at $r=0$ and $r=\frac{4\kappa
\mathcal{M}l^{2}}{\mu _{E}^2l^{2}+4}>0$. Also, $\frac{4\kappa
\mathcal{M}l^{2}}{\mu _{E}^2l^{2}+4}<-\frac{4\kappa l^{2}\mathcal{M}}{\mu
_{E}^2l^{2}-4}$. So, there is a horizon at $r=\frac{4\kappa \mathcal{M}%
l^{2}}{\mu _{E}^2l^{2}+4}$ and it is in the normal region, where $\partial _{\varphi}$ is spacelike and the solution represent a G\"{o}del Black Hole,~\cite{Banados}.
\end{enumerate}

A G\"{o}del black hole describes a black hole in a rotating G\"{o}del
background (with horizons in the normal region), while a G\"{o}del particle
describes a particle-like solution in a rotating G\"{o}del background. The
asymptotic behavior of these solutions is similar to the asymptotic behavior
of the G\"{o}del Universe (i.e. they are asymptotically G\"{o}del). \footnote{Notice
that by 'particle-like solution' we mean a solution which exhibits a naked
conical singularity; it is horizonless in the normal region. This is totally
analogous as when Deser-Jackiw-t'Hooft identified conical singularities in
3D gravity as particle-like solutions in GR.}

As pointed out in Ref.~\cite{Clement2} the gravitational constant can be
positive or negative in $2+1$ dimensions~\cite{jackiw}, also it is well know
that in topologically massive gravity the gravitational constant should be
taken as negative to avoid the appearance of ghosts,~\cite{deser}. We mention
that in this case we have $M>0$
and choose $\kappa <0$  in Eq.~(\ref{potential2}), to arrive to a real solution.

It is worth noticing that is possible to make the following coordinate
transformation $\phi \rightarrow i\phi $, $t\rightarrow it$ and $%
r\rightarrow -r$~\cite{Banados} to find a new solution, in this case the
metric and the gauge field become\footnote{Because $N$ is arbitrary we also make the
change $N\rightarrow -N=2$.}
\begin{equation}
ds^{2}=\frac{\psi (r)}{-f(r)}dt^{2}+\left( -f(r)\right) \left( d\phi +\frac{%
Cr}{-f(r)}dt\right) ^{2}-\frac{1}{\psi (r)}dr^{2}~,
\end{equation}%
\begin{equation}
A=\frac{\frac{\mu _{E}}{\kappa }}{\sqrt{-\frac{1}{\kappa }M}}dt+\left( \frac{%
\frac{1}{\kappa }}{\sqrt{-\frac{1}{\kappa }M}}+Hr\right) d\phi ~,
\end{equation}%
and
\begin{equation}
\mathcal{Q}=0~,\quad \mathcal{M}=-\frac{\left( \mu _{E}^{2}+4\Lambda
_{eff}\right) }{2\kappa }\frac{1}{M}~,\quad \mathcal{J}=0~.
\end{equation}%
In this case, we have a real solution for $M<0$ and $\kappa >0$ and we see
that our black hole generalizes the Einstein-Maxwell-Chern-Simons solution (with
negative cosmological constant) obtained in Ref.~\cite{Banados}\footnote{here the gravitational constant is positive.} to spacetime with torsion and
the solution admits naked closed timelike curves \footnote{In Ref.~\cite{Banados}, $\alpha =-\frac{\mu _{E}}{2}$.}.

We also find that when $\mu _{E}^{2}l^{2}=4$ the fluid disappears, the
energy-momentum tensor vanishes and the metric reduces to the BTZ metric in
Einstein-Cartan spacetime (also valid for the general solution, Eqs.~(\ref{solution1},~\ref{solution2})).

In the case of not charged solutions, from the gravitational Eqs.~(\ref{ec1}~-~\ref{ec5}) with $H=0$, we recover the BTZ black hole
in Riemann-Cartan spacetime $(M=0,~C=\frac{2}{l})$.

\subsection{Case: $\alpha _{3}\alpha
_{4}-\left( \frac{1}{2\kappa }\right) ^{2}=0$.}
In this case the field equations degenerate to a single equation, which is
\begin{equation}\label{degenerateeq}
R^{a}+\frac{1}{2\kappa \alpha _{3}}T^{a}+\frac{1}{2\left( 2\kappa \alpha
_{3}\right) ^{2}}\epsilon _{bc}^{a}e^{b}e^{c}=0~,
\end{equation}
making $2\kappa=1$ and by identifying, $\alpha _{3}=\frac{1}{\mu }$ and $\Lambda =-\frac{1}{l^{2}}$ and along with the condition $\alpha _{3}=-\frac{1}{\sqrt{-\Lambda }}$, yields to a point analogous to the chiral point, $\mu l=-1$, of TMG. Therefore, we consider the metric\footnote{For k=0, this metric corresponds to the metric of BTZ extremal.}
\begin{equation}
ds^{2}=f\left( r\right) ^{2}dt^{2}-\frac{1}{g\left( r\right) ^{2}}
dr^{2}-h\left( r\right) ^{2}\left( d\phi +C\left( r\right) dt\right)^{2}~,
\end{equation}
with
\begin{equation}
f\left( r\right) =\frac{r^{2}}{lh\left(r\right)}~,\quad g\left( r\right) =\frac{r}{l}~,\quad
C\left( r\right) =-\frac{kl\ln \left( r^{2}/r_{0}^{2}\right) }{h\left(
r\right) ^{2}}~,\quad h\left( r\right) =\sqrt{r^{2}+kl^{2}\ln \left( r^{2}/r_{0}^{2}\right)}~.
\end{equation}
Where, through simplicity and without the loss of generality we have chosen $M=0$, in Eq.~(5) of Ref.~\cite{Garbarz:2008qn}.
This metric, corresponds to a one-parameter deformation of GR solutions and is continuously connected
to the extremal BTZ black hole.
The vielbein is
\begin{equation}
e^{0}=f\left( r\right) dt~,
\end{equation}
\begin{equation}
e^{1}=\frac{1}{g\left( r\right) }dr~,
\end{equation}
\begin{equation}
e^{2}=h\left( r\right) \left( d\phi +C\left( r\right) dt\right)~,
\end{equation}
and the spin connection is given by Eq.~(\ref{conecction}), where the Riemannian connection is
\begin{equation}
\bar{\omega}^{0}=\frac{k}{h\left( r\right) }dt-\frac{r^{2}+kl^{2}}{lh\left(
r\right)}d\phi~,
\end{equation}
\begin{equation}
\bar{\omega}^{1}=\frac{kl}{r}\frac{l-\ln \left( r^{2}/r_{0}^{2}\right) }{%
h\left( r\right) ^{2}}dr~,
\end{equation}
\begin{equation}
\bar{\omega}^{2}=\frac{1}{h\left( r\right) }\left( \left( k\left( 1-\ln
\left( r^{2}/r_{0}^{2}\right) \right) -\frac{r^{2}}{l^{2}}\right)
dt-kl\left( 1-\ln \left( r^{2}/r_{0}^{2}\right) \right) d\phi \right)~,
\end{equation}
and the Riemannian curvature is\footnote{These expressions are given in Ref. \cite{Blagojevic:2010jv}.}
\begin{equation}
\bar{R}^{0}=\frac{2k}{h\left( r\right) ^{2}}e^{0}e^{1}+\left( \frac{2k}{%
h\left( r\right) ^{2}}-\frac{1}{l^{2}}\right) e^{1}e^{2}~,
\end{equation}
\begin{equation}
\bar{R}^{1}=-\frac{1}{l^{2}}e^{0}e^{2}~,
\end{equation}
\begin{equation}
\bar{R}^{2}=\left( \frac{2k}{h\left( r\right) ^{2}}+\frac{1}{l^{2}}\right)
e^{0}e^{1}+\frac{2k}{h\left( r\right) ^{2}}e^{1}e^{2}~.
\end{equation}
Now, through following ansatz for the contortion
\begin{equation}
K^{0}=Te^{0}+\chi \left( r\right) \left( e^{0}-e^{2}\right)~,
\end{equation}
\begin{equation}
K^{1}=Te^{1}~,
\end{equation}
\begin{equation}
K^{2}=\chi \left( r\right) \left( e^{0}-e^{2}\right)+Te^{2},
\end{equation}
With $T$ constant. The gravitational equations can be written as
\begin{equation}\label{aux}
\bar{R}^{a}+\bar{D}K^{a}+\frac{1}{2}\epsilon _{bc}^{a}K^{b}K^{c}+\frac{1}{%
2\kappa \alpha _{3}}\epsilon _{bc}^{a}K^{b}e^{c}+\frac{1}{2\left( 2\kappa
\alpha _{3}\right) ^{2}}\epsilon _{bc}^{a}e^{b}e^{c}=0~,
\end{equation}
and the solution to this equation is given by the following cases:
\begin{equation}
T=0,
\end{equation}
\begin{equation}
\chi \left( r\right) =-\frac{kl}{r^2+kl^2ln\left( \frac{r^2}{r^2_0}\right)}+\frac{Cte*r^2}{r^2+kl^2ln\left( \frac{r^2}{r^2_0}\right) }~,
\end{equation}
or
\begin{equation}
T=\frac{2}{l},
\end{equation}
\begin{equation}
\chi \left( r\right) =\frac{kl ln(r^{2})}{r^2+kl^2ln\left(\frac{r^2}{r^2_0}\right)}+\frac{Cte}{r^2+kl^2ln\left( \frac{r^2}{r^2_0}\right)}~,
\end{equation}
where, $Cte$ is an arbitrary integration constant that is consequence of the degeneracy of the field equations.
For $M\neq 0$ the solutions take the form
\begin{equation}
\chi \left( r\right) =-\frac{kl}{r^2+kl^2ln\left( \frac{r^2-4GMl^{2}}{r^2_0}\right)}+\frac{Cte*(r^2-4GMl^{2})}{r^2+kl^2ln\left( \frac{r^2-4GMl^{2}}{r^2_0}\right) }~,
\end{equation}
and
\begin{equation}\label{chi}
\chi \left( r\right) =\frac{kl ln(r^{2}-4GMl^{2})}{r^2+kl^2ln\left( \frac{r^2-4GMl^{2}}{r^2_0}\right)}+\frac{Cte}{r^2+kl^2ln\left( \frac{r^2-4GMl^{2}}{r^2_0}\right)}~,
\end{equation}
Therefore, in the last case the torsion is given by
\begin{equation}
T^0=\frac{4}{l}e^1e^2-\chi (r)e^0e^1-\chi (r)e^1e^2~,\quad
T^1=\frac{4}{l}e^0e^2,\quad
T^2=-\frac{4}{l}e^0e^1-\chi (r)e^0e^1-\chi (r)e^1e^2~.
\end{equation}
Note that the torsion is a function of $r$. Now, if we take the limit $k=0$ we recover the extremal BTZ metric with non constant torsion (if, in addition $Cte=0$, we recover BTZ metric with torsion).
As a result, we obtain for the conserved charges (applying directly expressions (3.3a) and (3.3b) of ~\cite{Blagojevic:2010jv} and taking the reference configuration $k=0$ and $M=0$) the following values for the mass and the angular momentum in this last case~Eq.(\ref{chi}):
\begin{equation}
\mathcal{M}l=-\mathcal{J}=\frac{kl}{2G}.
\end{equation}
This result gives the same conserved charges of the torsionless logarithmic solution of TMG. Note that these charges do not depend on the integration constant $Cte$.
Therefore, the logarithmic solution of TMG, found in Ref.~\cite{Garbarz:2008qn}, is also a solution of the Mielke-Baekler model.

The fact that solutions with logarithmic asymptotic behavior arise in the
degenerate point of Mielke-Baekler theory is reminiscent of what happens in
topologically massive gravity (TMG) at the chiral point, where such a
logarithmic behavior has been observed and leads to the so-called
Log-gravity. As pointed out in \cite{Santamaria:2011cz}, there exists a
remarkable similarity (though not equivalence) between the chiral point of
TMG and the degenerate point of Mielke-Baekler Chern-Simons gravity; the
fact we observe such a behavior to emerge at this special point of the space
of parameters supports this parallelism between both models.

We note that for this solution to emerge at the singular point requires $%
\alpha _{4}=-\frac{1}{l}\neq 0$. This implies that the translational
Chern-Simons term must be included in the action for this solution appears
in the case of adding torsion.

The logarithmic solution emerges in this theory because as we see from Eq.~(\ref{aux})  torsional degrees of freedom are playing a role analogous to that of
Cotton tensor of TMG and has not constant curvature neither constant torsion.
Both quantities depend on the radial coordinate $r$ and the sum $R^{a}+\frac{1}{2k\alpha _{3}}T^{a}$ is constant. Which, can be see
directly from Eq.~(\ref{degenerateeq}).

\section{Final remarks}

In this work, we have discussed black holes solutions to
three-dimensional Einstein theory with torsion coupled to topologically
massive gravity and charged under topologically massive electrodynamics. For vanishing torsion, the solution that we found by means a coordinate transformation can
be written as a three-dimensional Chern-Simons black holes. Also, we showed that the logarithmic solution of TMG is a solution of the Mielke-Baekler model at a special point of the space of parameters. Our
solution generalizes previous solutions reported in the literature, so that
it may contribute to a better understanding of the solutions content of this
interesting toy model of gravity. However, some of our particular cases that
we have analyzed are restricted at integrability conditions of the conserved
charges (See appendix C). So, it might be interesting to use another approach for the calculation of the conserved charges for this type of spacetime and compare them with the Hamiltonian method used here, e.g. see~\cite{Banados,Blagojevic:2010jv,Miskovic:2009kr}. Currently, we are working on the latter, also we are
determining some properties of this spacetime, which we expect to report in the near
future.

\section*{Acknowledgements}

The authors thank to B. Cvetkovic and J. Saavedra
for such enlightening comments. Specially, we would like to thank G. Giribet and Olivera Miskovic for helpful suggestions and comments.
Y.V. was supported by Direcci\'{o}n de Investigaci\'{o}n y Desarrollo, Universidad de la Frontera, DIUFRO DI11-0071 and partially support by ``Programas de Apoyo a la Inserci\'{o}n DI10-600''.

\appendix{}

\section{Gravitational equations}
In this appendix we write a number of auxiliary expressions used in the text. We start with covariant derivatives of the vector $s^{a}$ related to the energy-momentum current
\begin{equation}
\bar{D}s^{a}=ds^{a}+\epsilon _{bc}^{a}\bar{\omega}^{b}s^{c}~,
\end{equation}%
Its components are
\begin{gather}
\nonumber \bar{D}s^{0}=-\left( \frac{H^{2}}{4f^{2}}f^{\prime }\left( 3\psi
+C^{2}r^{2}\right) \sqrt{\psi } +\frac{H^{2}}{4f}\left( 3\psi ^{\prime
}+2C^{2}r\right) \sqrt{\psi }\right) e^{0}e^{1} \\
\nonumber -\frac{H^{2}}{8\sqrt{\psi }}\left( 3\psi +C^{2}r^{2}\right) \left( \frac{%
\psi ^{\prime }f-\psi f^{\prime }}{f^{2}}\right) e^{0}e^{1} +CH^{2}\left(
\frac{\psi }{f}+\frac{r}{2f}\psi ^{\prime }-\frac{r\psi }{2f^{2}}f^{\prime
}\right) e^{1}e^{2} \\
\nonumber -\frac{C^{2}H^{2}r\sqrt{\psi }}{f}\left( 1-r\frac{f^{\prime }}{f}\right)
e^{0}e^{1}-\frac{H^{2}}{8f\sqrt{\psi }}\left( \psi ^{\prime }-\psi \frac{%
f^{\prime }}{f}\right) \left( C^{2}r^{2}-\psi \right) e^{0}e^{1} \\
-\frac{1}{2}\frac{C^{3}H^{2}r^{2}}{f}\left( 1-r\frac{f^{\prime }}{f}\right)
e^{1}e^{2}-\frac{1}{2}\frac{C^{2}H^{2}r\sqrt{\psi }}{f}\left( 1-r\frac{%
f^{\prime }}{f}\right) e^{0}e^{1}~,
\end{gather}%
\begin{equation}
\bar{D}s^{1}=-\frac{H^{2}}{f}\left( \frac{1}{2}C\left( 1-r\frac{f^{\prime }}{%
f}\right) \left( \psi +C^{2}r^{2}\right) -Cr\left( \frac{\psi ^{\prime }}{2}%
-\psi \frac{f^{\prime }}{f}\right) \right) e^{0}e^{2}~,
\end{equation}%
\begin{gather}
\nonumber\bar{D}s^{2}=\frac{H^{2}}{4}\left( -\frac{f^{\prime }}{2f^{2}}\left( \psi
+3C^{2}r^{2}\right) +\frac{1}{f}\left( \psi ^{\prime }+6C^{2}r\right)
\right) \sqrt{\psi }e^{1}e^{2} \\
\nonumber-\frac{CH^{2}}{4f}\left( \psi +3C^{2}r^{2}\right) \left( 1-r\frac{f^{\prime }%
}{f}\right) e^{0}e^{1} +CH^{2}\left( \frac{\psi }{f}+\frac{r\psi ^{\prime }}{%
f}-\frac{3}{2}\frac{r\psi }{f^{2}}f^{\prime }\right) e^{0}e^{1} \\
\nonumber-\frac{1}{4}\frac{CH^{2}}{f}\left( \psi +C^{2}r^{2}\right) \left( 1-r\frac{%
f^{\prime }}{f}\right) e^{0}e^{1} -\frac{1}{2}\frac{C^{2}H^{2}r}{f}\sqrt{%
\psi }\left( 1-r\frac{f^{\prime }}{f}\right) e^{1}e^{2} \\
+\frac{1}{8}H^{2}\sqrt{\psi }\frac{f^{\prime }}{f^{2}}\left( C^{2}r^{2}-\psi
\right) e^{1}e^{2}~.
\end{gather}%
Next, we need the following vector constructed from $s^a$
\begin{equation}
X^{a}=\epsilon _{bc}^{a}s^{b}s^{c}~,
\end{equation}%
Its components read
\begin{equation}
X^{0}=-\frac{1}{8}\left( E^{2}-B^{2}\right) \left( 3E^{2}+B^{2}\right)
e^{1}e^{2}+\frac{1}{2}EB\left( E^{2}-B^{2}\right) e^{0}e^{1}~,
\end{equation}%
\begin{equation}
X^{1}=-\left( \frac{1}{8}\left( 3E^{2}+B^{2}\right) \left(
E^{2}+3B^{2}\right) -2E^{2}B^{2}\right) e^{0}e^{2}~,
\end{equation}%
\begin{equation}
X^{2}=-\frac{1}{8}\left( E^{2}-B^{2}\right) \left( E^{2}+3B^{2}\right)
e^{0}e^{1}+\frac{1}{2}EB\left( E^{2}-B^{2}\right) e^{1}e^{2}~,
\end{equation}%
and the vector
\begin{equation}
Y^{a}=\epsilon _{bc}^{a}s^{b}e^{c}~,
\end{equation}
which components are
\begin{equation}
Y^{0}=\frac{1}{2}\left( E^{2}+B^{2}\right) e^{1}e^{2}-EBe^{0}e^{1}~,
\end{equation}%
\begin{equation}
Y^{1}=-\frac{1}{2}\left( B^{2}-E^{2}\right) e^{0}e^{2}~,
\end{equation}%
\begin{equation}
Y^{2}=\frac{1}{2}\left( E^{2}+B^{2}\right) e^{0}e^{1}-EBe^{1}e^{2}~.
\end{equation}
Using the useful expressions from above, the gravitational equations can be written as the following algebraic
system of equations
\begin{eqnarray}
Cr\frac{f^{\prime \prime }}{f} &=&\gamma \frac{CH^{2}r}{f}-m\frac{H^{2}}{%
4f^{2}}f^{\prime }\left( 3\psi +C^{2}r^{2}\right) +m\frac{H^{2}}{4f}\left(
3\psi ^{\prime }+2C^{2}r\right) +m\frac{1}{2}H^{2}\left( \frac{\psi ^{\prime
}f-\psi f^{\prime }}{f^{2}}\right)  \notag \\
&&-\frac{3}{2}m\frac{C^{2}H^{2}r}{f}\left( 1-r\frac{f^{\prime }}{f}\right) -%
\frac{m^{2}}{8}\frac{CH^{4}r}{f^{2}}\left( C^{2}r^{2}-\psi \right) ~,
\label{ec1}
\end{eqnarray}%
\begin{gather}
\frac{1}{4}\psi ^{\prime }\frac{f^{\prime }}{f}+\frac{1}{2}\psi \frac{%
f^{\prime \prime }}{f}-\frac{1}{4}\psi \frac{f^{\prime 2}}{f^{2}}+\frac{1}{4}%
C^{2}\left( 1-r\frac{f^{\prime }}{f}\right) ^{2}=\frac{1}{4}\gamma \frac{%
H^{2}}{f}\left( \psi +C^{2}r^{2}\right)  \notag \\
+\frac{m}{2}CH^{2}\left( \frac{\psi }{f}+\frac{r\psi ^{\prime }}{2f}-\frac{%
r\psi }{2f^{2}}f^{\prime }\right)-\frac{m}{4}\frac{C^{3}H^{2}r^{2}}{f}\left(
1-r\frac{f^{\prime }}{f}\right)  \notag \\
-\frac{m^{2}}{64}\frac{H^{4}}{f^{2}}\left( \psi +3C^{2}r^{2}\right) \left(
C^{2}r^{2}-\psi \right) -\Lambda _{eff}~,  \label{ec2}
\end{gather}%
\begin{gather}
-\left( \frac{1}{4}\frac{f^{\prime }}{f}\left( \psi ^{\prime }-\psi \frac{%
f^{\prime }}{f}\right) +\frac{1}{4}C^{2}\left( 1-r\frac{f^{\prime }}{f}%
\right) ^{2}\right) =\frac{1}{4}\gamma \frac{H^{2}}{f}\left( \psi
-C^{2}r^{2}\right)  \notag \\
+\frac{m}{4}\frac{CH^{2}}{f}\left( 1-r\frac{f^{\prime }}{f}\right) \left(
\psi +C^{2}r^{2}\right)-\frac{m}{2}\frac{CH^{2}r}{f}\left( \frac{\psi
^{\prime }}{2}-\psi \frac{f^{\prime }}{f}\right)  \notag \\
+\frac{m^{2}}{8}\left( \frac{1}{8}\frac{H^{4}}{f^{2}}\left( 3\psi
+C^{2}r^{2}\right) \left( \psi +3C^{2}r^{2}\right) -\frac{%
2C^{2}H^{4}r^{2}\psi }{f^{2}}\right) +\Lambda _{eff}~,  \label{ec3}
\end{gather}%
\begin{gather}
-\frac{1}{2}\psi \frac{f^{\prime \prime }}{f}+\frac{1}{2}\psi ^{\prime
\prime }-\frac{3}{4}\psi ^{\prime }\frac{f^{\prime }}{f}+\frac{3}{4}\psi
\frac{f^{\prime 2}}{f^{2}}-\frac{3}{4}C^{2}\left( 1-r\frac{f^{\prime }}{f}%
\right) ^{2}=-\frac{1}{4}\gamma \frac{H^{2}}{f}\left( \psi +C^{2}r^{2}\right)
\notag \\
+m\frac{CH^{2}}{4f}\left( 1-r\frac{f^{\prime }}{f}\right) \left( \psi
+2C^{2}r^{2}\right) -\frac{m}{2}\frac{CH^{2}}{f}\left( \psi +r\psi ^{\prime
}-\frac{3}{2}r\psi \frac{f^{\prime }}{f}\right)  \notag \\
+\frac{m^{2}}{64}\frac{H^{4}}{f^{2}}\left( -\psi +C^{2}r^{2}\right) \left(
3\psi +C^{2}r^{2}\right) -\Lambda _{eff}~,  \label{ec4}
\end{gather}%
\begin{gather}
-Cr\frac{f^{\prime \prime }}{f}=-\gamma \frac{CH^{2}r}{f}-m\frac{H^{2}}{4}%
\left( -\frac{f^{\prime }}{2f^{2}}\left( \psi +3C^{2}r^{2}\right) +\frac{1}{f%
}\left( \psi ^{\prime }+6C^{2}r\right) \right)  \notag \\
+\frac{1}{2}m\frac{C^{2}H^{2}r}{f}\left( 1-r\frac{f^{\prime }}{f}\right) -m%
\frac{H^{2}}{8}\frac{f^{\prime }}{f^{2}}\left( -\psi +C^{2}r^{2}\right) +%
\frac{m^{2}}{8}\frac{CH^{4}r}{f^{2}}\left( -\psi +C^{2}r^{2}\right)~.
\label{ec5}
\end{gather}
By adding Eq.~(\ref{ec1}) and Eq.~(\ref{ec5}), we obtain
\begin{equation}
-f^{\prime }\psi +C^{2}r^{2}f^{\prime }-2C^{2}rf+f\psi ^{\prime }=0~,
\end{equation}
from this equation we find the following constraints between the
coefficients
\begin{equation}
P=\frac{R}{L}M+C^{2}~,
\end{equation}%
\begin{equation}
Q=\frac{R}{L}N~.
\end{equation}
Therefore, the function $\psi (r)$ forms as
\begin{equation}
\psi (r)=\frac{R}{L}f(r)+C^{2}r^{2}~.  \label{phi}
\end{equation}

Now, as we go back to the system of equations and find that only two equations
are independent allowing us to get the constants $M$ and $H$. Adding
Eq.~(\ref{ec2}) and Eq.~(\ref{ec3}), yields
\begin{equation}
\frac{m^{2}}{8}\frac{R}{L}H^{4}+\left( \gamma +\frac{3}{2}mC\right)
H^{2}-2M=0~.
\end{equation}%
By multiplying equation Eq.~(\ref{ec2}) by 3 and adding the result to
Eq.~(\ref{ec4}), we obtain
\begin{gather}
\psi \frac{f^{\prime \prime }}{f}+\frac{1}{2}\psi ^{\prime \prime }=\frac{
\gamma }{2}\frac{H^{2}}{f}\left( \psi +C^{2}r^{2}\right) +\frac{5}{4}mCH^{2}
\frac{\psi }{f}+\frac{1}{4}m\frac{CH^{2}r}{f}\psi ^{\prime }-\frac{1}{4}m%
\frac{CH^{2}rf^{\prime }\psi }{f^{2}}  \notag \\
-\frac{1}{4}m\frac{C^{3}H^{2}r^{2}}{f}+\frac{1}{4}m\frac{C^{3}H^{2}r^{3}f^{
\prime }}{f^{2}}+\frac{m^{2}}{8}\frac{H^{4}}{f^{2}}C^{2}r^{2}\left( \psi
-C^{2}r^{2}\right) -4\Lambda _{eff}~,
\end{gather}
and from this equation we arrive to another independent relation
\begin{equation}
3M\frac{R}{L}+C^{2}+4\Lambda _{eff}=\frac{1}{2}\left( \gamma +\frac{5}{2}
mC\right) H^{2}\frac{R}{L}~.
\end{equation}
It is worth noting that when $L=0$, $R$ must vanishes, and our solution is given by replacing
$R/L$ by $Q/N$.

\section{Asymptotic behavior}

The asymptotic behavior of the vielbein and the Cartan connection are
determined by the following expansion
\begin{equation}
e_{t}^{0}\thicksim \sqrt{\frac{P}{M}}\left( 1+\frac{1}{2}\left( \frac{Q}{P}-%
\frac{N}{M}\right) \frac{1}{r}+\mathcal{O}(r^{2})\right)~,
\end{equation}
\begin{equation}
e_{r}^{1}\backsim \mathcal{O}(r)~,
\end{equation}
\begin{equation}
e_{\phi }^{2}\thicksim \sqrt{M}\left( r+\frac{1}{2}\frac{N}{M}+\frac{1}{2}%
\left( \frac{L}{M}-\frac{1}{4}\frac{N^{2}}{M^{2}}\right) \frac{1}{r}+%
\mathcal{O}(r^{2})\right)~,
\end{equation}
\begin{equation}
e_{t}^{2}\backsim \frac{C}{\sqrt{M}}\left( 1-\frac{1}{2}\frac{N}{M}\frac{1}{r%
}+\mathcal{O}(r^{2})\right)~.
\end{equation}
\begin{equation}
\bar{\omega}_{t}^{0}\backsim -\frac{1}{2}C\sqrt{\frac{P}{M}}\left( 1+\frac{1%
}{2}\left( \frac{Q}{P}-\frac{N}{M}\right) \frac{1}{r}+\mathcal{O}%
(r^{2})\right)~,
\end{equation}
\begin{equation}
\bar{\omega}_{\phi }^{0}\backsim -\sqrt{PM}\left( r+\frac{1}{2}\frac{Q}{P}+%
\frac{1}{2}\left( -\frac{L}{M}+\frac{R}{P}+\frac{1}{8}\frac{N^{2}}{M^{2}}-%
\frac{1}{4}\frac{Q^{2}}{P^{2}}\right) \frac{1}{r}+\mathcal{O}(r^{2})\right)~,
\end{equation}
\begin{equation}
\bar{\omega}_{r}^{1}\backsim \mathcal{O}(r)~,
\end{equation}
\begin{equation}
\bar{\omega}_{t}^{2}\thicksim \frac{1}{2}\frac{C^{2}}{\sqrt{M}}\left( -1+%
\frac{1}{2}\frac{N}{M}\frac{1}{r}+\mathcal{O}(r^{2})\right)~,
\end{equation}
\begin{equation}
\bar{\omega}_{\phi }^{2}\backsim \frac{1}{2}C\sqrt{M}\left( -r+\frac{1}{2}%
\frac{N}{M}+\frac{3}{2}\left( \frac{L}{M}-\frac{1}{4}\frac{N^{2}}{M^{2}}%
\right) \frac{1}{r}+\mathcal{O}(r^{2})\right)~.
\end{equation}
\begin{equation}
K_{t}^{0}\thicksim \frac{1}{2}\left( a-\frac{3}{4}mH^{2}\frac{R}{L}\right)
\sqrt{\frac{P}{M}}\left( 1+\frac{1}{2}\left( \frac{Q}{P}-\frac{N}{M}\right)
\frac{1}{r}+\mathcal{O}(r^{2})\right)~,
\end{equation}
\begin{gather}
\nonumber K_{\phi }^{0}\thicksim \frac{1}{2}mCH^{2}\sqrt{\frac{P}{M}}\left( r+\frac{1}{%
2}\left( \frac{Q}{P}-\frac{N}{M}\right) +\frac{1}{2}\left( \frac{R}{P}-\frac{%
L}{M}-\frac{1}{4}\frac{Q^{2}}{P^{2}}-\frac{1}{2}\frac{QN}{PM}+\frac{3}{4}%
\frac{N^{2}}{M^{2}}\right) \frac{1}{r}\right) \\
+\mathcal{O}(r^{2})~,
\end{gather}
\begin{equation}
K_{t}^{2}\thicksim \frac{1}{2}\left( a\frac{C}{\sqrt{M}}-\frac{3}{4}mH^{2}%
\frac{R}{L}\frac{C}{\sqrt{M}}\right) \left( 1-\frac{N}{2M}\frac{1}{r}+%
\mathcal{O}(r^{2})\right)~,
\end{equation}
\begin{gather}
\nonumber K_{\phi }^{2}\thicksim \frac{1}{2}a\sqrt{M}\left( r+\frac{1}{2}\frac{N}{M}+%
\frac{1}{2}\left( \frac{L}{M}-\frac{1}{4}\frac{N^{2}}{M^{2}}\right) \frac{1}{%
r}\right) +\frac{1}{8}m\frac{H^{2}}{\sqrt{M}} \left( P+3C^{2}\right)r \\
\nonumber +\frac{1}{8}m\frac{H^{2}}{\sqrt{M}}\left( Q-\frac{1}{2}\frac{N}{M}\left(
P+3C^{2}\right)+\left( R-\frac{1}{2}\frac{NQ}{M}+\left( -\frac{1}{2}\frac{L}{%
M}+\frac{3}{8}\frac{N^{2}}{M^{2}}\right) \left( P+3C^{2}\right) \right)
\frac{1}{r}\right) \\
+\mathcal{O}(r^{2})~.
\end{gather}
\begin{equation}
\omega _{\mu }^{a}=\bar{\omega}_{\mu }^{a}+K_{\mu }^{a}~,
\end{equation}
where $\mathcal{O}(r^{n})\sim \frac{1}{r^{n}}$.

\section{Conserved charges}
In this appendix we calculate the mass, angular momentum and electric charge
of our solution. The asymptotic behavior of the vielbein and the Cartan
connection is given in appendix B. The canonical generator has the following
form in the asymptotic region
\begin{equation}
G=-G_{1}-G_{3}~,
\end{equation}%
with
\begin{equation}
G_{1}=\xi ^{\rho }\left[ e_{\rho }^{a}\mathcal{H}_{a}+\omega _{\rho }^{a}%
\mathcal{K}_{a}+\left( \partial _{\rho }e_{t}^{a}\right) \pi _{a}^{t}+\left(
\partial _{\rho }\omega _{t}^{a}\right) \Pi _{t}^{a}+\left( \partial _{\rho
}A_{t}\right) \pi ^{t}\right] ~,
\end{equation}
\begin{equation}
G_{3}=-\lambda \left( \partial _{\alpha }\pi ^{\alpha }-2\mu _{E}\epsilon
^{t\alpha \beta }\partial _{\alpha }A_{\beta }\right) ~,
\end{equation}
\begin{equation}
\pi ^{\alpha }=-kF^{t\alpha }-\mu _{E}\epsilon ^{t\alpha \beta }A_{\beta }~.
\end{equation}
The expressions for $\mathcal{H}_{a}$ and $\mathcal{K}_{a}$ can be found in
appendix C of~\cite{Blagojevic} (see also~\cite{Blagojevic5}). The variation
of the generator produces
\begin{eqnarray}
\nonumber \delta G_{1} &=&\xi ^{\rho }\left( -2\epsilon ^{t\alpha \beta }\partial
_{\alpha }\left[ e_{\rho }^{a}\delta \left( \frac{1}{2\kappa }\omega
_{a\beta }+\alpha _{4}e_{a\beta }\right) +\omega _{\rho }^{a}\delta \left(
\frac{1}{2\kappa }e_{a\beta }+\alpha _{3}\omega _{a\beta }\right) \right]
+\delta \tau _{\rho }^{t}\right) \\
&&+~regular~terms~,
\end{eqnarray}%
\begin{equation}
\delta G_{3}=-\lambda \left( \partial _{\alpha }\delta \pi ^{\alpha }-2\mu
_{E}\epsilon ^{t\alpha \beta }\partial _{\alpha }\delta A_{\beta }\right) ~,
\end{equation}%
where, $\delta \tau _{\rho }^{t}=\frac{1}{2}\epsilon ^{t\alpha \beta
}e_{\rho }^{i}\delta \chi _{i\alpha \beta }$ comes from the Maxwell
energy-momentum current which is given by\footnote{%
See Appendix C ~\cite{Blagojevic}.} $\chi ^{i}=\frac{1}{2}\chi _{\mu \nu
}^{i}dx^{\mu }dx^{\nu }$. Using the asymptotic behavior of the vielbein and
the connection (Appendix B) we obtain the conserved charges. For simplicity,
we take $\alpha _{3}=0$. The integrability conditions of our solution is
satisfied only for the case that $L$ depends of $M$ $\left( L=f(M)\right) $, e.g. for $L=\frac{1}{M}$ we obtain
\begin{equation}
\mathcal{Q}=0~
\end{equation}
\begin{eqnarray}
\nonumber\mathcal{M} &=&-\frac{1}{M}\left( \frac{1}{2\kappa }\rho +\frac{1}{\kappa }a\mu _{E}-%
\frac{1}{4\kappa }m\mu _{E}\sigma +2\alpha _{4}\mu _{E}+ \frac{1}{2}\rho \sqrt{\frac{\sigma }{\rho }}\frac{A_{t}\sqrt{M}}{\mu _{E}}\right)\\
&&-\frac{\alpha _{3}}{M}\left(
\frac{\rho }{2}+\frac{\mu _{E}^{2}}{2}+\frac{1}{2}a\mu _{E}+\frac{m\sigma
\mu _{E}}{8}\right) \left( -\mu _{E}+a-\frac{3}{4}m\sigma \right)~,
\end{eqnarray}
\begin{equation}
\mathcal{J}=-\left[ \frac{1}{2\kappa }\left( a+\mu _{E}+\frac{1}{4}m\sigma
\right) +\frac{\alpha _{3}}{4}\left( a+\mu _{E}+\frac{1}{4}m\sigma \right)
^{2}+\alpha _{4}+\alpha _{3}\rho -\frac{1}{2}\sqrt{\frac{\sigma }{\rho }}%
A_{t}\sqrt{M}\right] \frac{1}{M}~,
\end{equation}
and
\begin{gather}
\nonumber
A_{t}\sqrt{M}= 2{\sqrt{\frac{\rho }{\sigma }}}\left[\frac{1}{\kappa }\left( \frac{a}{2}+\frac{m}{8}\frac{\sigma }{\rho }\left(
\rho +4\mu _{E}^{2}\right) -\frac{\mu _{E}}{2}\right)
+\alpha _{4}-\mu _{E}\frac{\sigma }{\rho}\right]\\
2\alpha_{3}{\sqrt{\frac{\rho }{\sigma }}}\left[
-\left( \rho +\mu _{E}^{2}\right)\left(-1+\frac{1}{2}m\mu _{E}\frac{\sigma}{\rho}\right)^{2}+\left(\frac{a}{2}+\frac{m}{8}\frac{\sigma}{\rho}\left(\rho+4\mu _{E}^{2}\right)-\frac{\mu _{E}}{2}\right)^{2}
\right]~,
\end{gather}
\begin{equation}
A_{\phi }=I+Hr=\frac{A_{t}}{\mu _{E}}+Hr~,
\end{equation}
where $\mathcal{Q}$, $\mathcal{M}$ and $\mathcal{J}$ denote the electric charge,
the mass and the angular momentum, respectively.

In the limit $\alpha _{3}\rightarrow 0$ ($m\rightarrow 0$) the equations (%
\ref{ecuacion1}) and (\ref{ecuacion2}) can be written as
\begin{equation}
-\kappa H^{2}=M~,
\end{equation}%
\begin{equation}
3M\frac{R}{L}+C^{2}+4\Lambda _{eff}=-\kappa H^{2}\frac{R}{L}~,
\end{equation}
and we obtain
\begin{equation}
H^{2}\frac{R}{L}=\frac{1}{2\kappa }\left( \mu _{E}^{2}+4\Lambda _{eff}\right)~,
\end{equation}
\begin{equation}
M\frac{R}{L}=-\frac{1}{2}\left( \mu _{E}^{2}+4\Lambda _{eff}\right)~.
\end{equation}
Also, the conserved charges are given by
\begin{equation}
\mathcal{Q}=0~,
\end{equation}
\begin{equation}
\mathcal{M}=\frac{\left( \mu _{E}^{2}+4\Lambda _{eff}\right) }{2\kappa }%
\frac{1}{M}~,
\end{equation}
\begin{equation}
\mathcal{J}=-\frac{\mu _{E}}{\kappa }\frac{1}{M}~,
\end{equation}
and
\begin{equation}
A=\frac{\frac{\mu _{E}}{\kappa }}{\sqrt{-\frac{1}{\kappa }M}}dt+\left( \frac{%
\frac{1}{\kappa }}{\sqrt{-\frac{1}{\kappa }M}}+Hr\right) d\phi~.
\label{potential}
\end{equation}
So, we get the following solution for the metric and gauge field
\begin{equation}
f(r)=\frac{1}{2\kappa l^{2}\mathcal{M}}\left( \mu _{E}^{2}l^{2}-4\right)
r^{2}+2r-\frac{\kappa \mathcal{J}}{\mu _{E}}~,
\end{equation}
\begin{equation}
\psi (r)=\frac{1}{2l^{2}}\left( \mu _{E}^{2}l^{2}+4\right) r^{2}-2\kappa
\mathcal{M}r+\frac{\kappa ^{2}\mathcal{MJ}}{\mu _{E}}~,
\end{equation}
\begin{equation}
A=\sqrt{\frac{1}{4\kappa ^{2}l^{2}\mathcal{M}}\left( \mu
_{E}^{2}l^{2}-4\right) }\left[ \frac{\frac{1}{\kappa }\mu _{E}}{\frac{1}{%
4\kappa ^{2}l^{2}\mathcal{M}}\left( \mu _{E}^{2}l^{2}-4\right) }dt+\left( r+%
\frac{\frac{1}{\kappa }}{\frac{1}{4\kappa ^{2}l^{2}\mathcal{M}}\left( \mu
_{E}^{2}l^{2}-4\right) }\right) d\phi \right]~.
\end{equation}
For $L=0$ and $\alpha _{3}=0$, we obtain
\begin{equation}
\mathcal{Q}=0~,\quad \mathcal{M}=\frac{\left( \mu _{E}^{2}+4\Lambda
_{eff}\right) }{2\kappa }\frac{1}{M}~,\quad \mathcal{J}=0~,
\end{equation}
\begin{equation}
A=\frac{\frac{\mu _{E}}{\kappa }}{\sqrt{-\frac{1}{\kappa }M}}dt+\left( \frac{%
\frac{1}{\kappa }}{\sqrt{-\frac{1}{\kappa }M}}+Hr\right) d\phi ~.
\end{equation}

\end{document}